# SMALL-SIGNAL CAPACITANCE AND CURRENT PARAMETER MODELING IN LARGE-SCALE HIGH-FREQUENCY GRAPHENE FIELD-EFFECT TRANSISTORS

G.I. Zebrev, A.A. Tselykovskiy, D.K. Batmanova, E.V. Melnik

Department of Micro- and Nanoelectronics, National Research Nuclear University MEPHI, 115409, Kashirskoe sh., 31, Moscow, Russia, gizebrev@mephi.ru

## Abstract

The analytical model of the small-signal current and capacitance characteristics of RF graphene FET is presented. The model is based on explicit distributions of chemical potential in graphene channels (including ambipolar conductivity at high source-drain bias) obtained in the framework of drift-diffusion current continuity equation solution. Small-signal transconductance and output conductance characteristics are modeled taking into account the two modes of drain current saturation including drift velocity saturation or electrostatic pinch-off. Analytical closed expression for the complex current gain and the cutoff frequency of high-frequency GFETs are obtained. The model allows describe an impact of parasitic resistances, capacitances, interface traps on extrinsic current gain and cut-off frequency.

## 1. Introduction

The radio frequency (RF) electronic devices play a central role in modern telecommunication systems. The new communication systems demand high frequency low power consumption with high degree of integration, along with good performance even under harsh environment such as radiation, etc. The unique properties of graphene [1] yield new opportunities to improve radio frequency low noise amplifiers. Graphene exhibits a very large carrier mobility, which is at least one order of magnitude greater than in Si that enables creation of devices with high current density and operational efficiency [2, 3]. The lack of a bandgap put a huge obstacle in applications of large-area graphene field-effects transistors (GFET) in digital circuits due to low on–off current ratios. Nevertheless the GFETs in RF applications are not required good on-off characteristics in itself and can benefit from the high mobility values offered by large-area graphene [4]. The significant progress in fabrication of RF GFETs with high performance characteristics (with frequencies as high as 300 GHz) have been reported [5].

One of the main possible application fields of graphene-based devices is space-borne RF telecommunications systems. Therefore, it is important to study the impact of radiation-induced interface traps on the high-frequency behavior of graphene RF transistors and, particularly, cutoff frequency. We have developed in this work a quantitative model of the capacitive and current small-signal parameters with consideration of interface trap buildup impact.

The paper is organized as follows. Sec.2 is devoted to the model background equations. Analytical I-V model in diffusion-drift approximation with two type of current saturation and a new unified approach to the current saturation mode modeling are briefly described in Sec.3. Capacitance and current small-signal parameters models are derived in Sec. 4-5. Frequency-dependent current gain and cutoff frequency simulations are presented in Sec.6.

## 2. Model background equations

This paper is based mainly on the physical model of GFET operation described in Ref.[6]. In this section we recall briefly the main equations of the diffusion-drift model for I-V characteristics in graphene presented in. Based on analytical solution of the current continuity equation in a diffusion-drift approximation we have obtained the explicit relationships for the distributions of the chemical and electrostatic potentials along the channel length separately and electrochemical potential as a whole

$$\varphi(x) - \varphi(0) = -\frac{\varepsilon_D}{\kappa e} \ln\left[1 - \frac{x}{L}\left[1 - \exp\left(-\frac{\kappa}{1+\kappa}\frac{eV_D}{\varepsilon_D}\right)\right]\right], \tag{1a}$$

$$\zeta(x) - \zeta(0) = \varepsilon_D \ln\left[1 - \frac{x}{L}\left[1 - \exp\left(-\frac{\kappa}{1+\kappa}\frac{eV_D}{\varepsilon_D}\right)\right]\right], \tag{1b}$$

$$\mu(x) = \mu(0) + \varepsilon_D \frac{1+\kappa}{\kappa} \ln\left[1 - \frac{x}{L}\left[1 - \exp\left(-\frac{\kappa}{1+\kappa}\frac{eV_D}{\varepsilon_D}\right)\right]\right], \tag{1c}$$

where $\zeta(0)$, $\mu(0)$ and $\varphi(0)$ are the chemical, electrochemical and electrostatic potentials nearby the source controlled by the gate-source bias $V_{GS}$, $\varepsilon_D = n_s/(dn_s/d\varepsilon_F)$ is the diffusion energy near the source, and the ratio of diffusion to drift current is expressed through the oxide ($C_{ox}$), the quantum ($C_Q$) and the interface trap ($C_{it}$) capacitances per unit area

$$\kappa = -\left(\frac{\partial \zeta}{e\partial \varphi}\right)_{V_G} = \frac{(\partial V_G/\partial \varphi)_\zeta}{e(\partial V_G/\partial \zeta)_\varphi} = \frac{C_{ox}}{C_Q + C_{it}}. \tag{2}$$

The total drain current at constant temperature can be written as gradient of the electrochemical potential taken in the vicinity of the source

$$I_D = -W\mu_0 n_s(0)\left(\frac{d\mu}{dx}\right)_{x=0} = e\frac{W}{L}D_0 N_s(0)\frac{1+\kappa}{\kappa}\left(1-\exp\left(-\frac{\kappa}{1+\kappa}\frac{eV_D}{\varepsilon_D}\right)\right) \equiv$$
$$\equiv I_{DSAT}\left(1-\exp\left(-2\frac{V_{DS}}{V_{DSAT}}\right)\right) \tag{3}$$

where $W$ is the channel width, $D_0$ is the diffusion constant and the Einstein relation $D_0 = \mu_0 \varepsilon_D/e$ is employed. Notice that the total two-dimensional charge density $eN_S = e(n_e + n_h) \cong en_S = e(n_e - n_h)$ practically equals to charge imbalance density excepting the vicinity of the charge neutrality point where diffusion-drift approximation is failed. The characteristic saturation source-drain voltage $V_{DSAT}$ can be defined as follows [6]

$$V_{DSAT} = 2\frac{1+\kappa}{\kappa}\frac{\varepsilon_D}{e} = V_G - V_{NP} + en_S(0)/C_{ox}, \tag{4}$$

where $\varepsilon_F = \zeta(0)$ is the Fermi energy (the same chemical potential) nearby the source (recall that $\varepsilon_D \cong \varepsilon_F/2$ far enough from the neutrality point in graphene). The chemical potential near the drain is expressed from Eq.1b as $\zeta(L) = (1 - V_D/V_{DSAT})\varepsilon_F$, and the condition $V_D = V_{DSAT}$ corresponds to zero of the chemical potential and current saturation due to electrostatic blocking which is known as pinch-off for silicon MOSFETs [7]. It is instructive to compare saturation voltages in GFETs and Si-MOSFETs where above the threshold voltage $V_T$ we have

$$V_{DSAT} = \varepsilon_F + \frac{en_S}{C_{ox} + C_D} \cong \frac{V_{GS} - V_T}{1+\xi}, \tag{5}$$

here $C_D$ is the depletion layer capacitance and $\xi = C_D/C_{ox}$ is the Si substrate influence factor. The channel capacitance is defined as

$$C_{CH} = e\left(\frac{\partial N_S}{\partial V_G}\right) = \frac{C_{ox}}{1 + \frac{C_{ox} + C_{it}}{C_Q}} \tag{6}$$

Recalling $en_S = C_Q \varepsilon_D$ and Eq.2 and 4 one can get useful relations

$$C_{CH} = \frac{\kappa C_Q}{1+\kappa} = \frac{2en_S}{V_{DSAT}}; \quad e(n_S - n_i) = C_{CH} V_{DSAT}/2. \tag{7}$$

The latter relation corrected at the CNP ($\varepsilon_F = 0$) in homogeneous graphene by residual concentration $n_i = (\pi/6)(k_B T/\hbar v_0)^2$ and connecting the analytic equations for channel capacitance, saturation voltage and charge density yields excellent numerical exactness for all temperatures and oxide gate parameters.

### 3. Saturation current DC parameters of GFET

#### A. Two modes of drain current saturation

The field-effect transistor is fundamentally non-linear device working at large biases generally on all electrodes. The saturation of the channel current in FETs at high source-drain electric field has two-fold origin, namely, (i) the current blocking due to carrier density depletion near the drain, and (ii) the carrier velocity saturation due to optical phonon emission. The saturation current for pinch-off case arising due to saturation of lateral electric field near the source is represented as follows

$$I_{DSAT} = \frac{W}{L} e D_0 n_S(0) \frac{1+\kappa}{\kappa} = \frac{W}{L} \frac{e n_S \sigma_0}{C_{CH}} = \frac{W}{L} \sigma_0(0) \frac{V_{DSAT}}{2}, \qquad (8)$$

where the Einstein relation connecting low-field conductivity $\sigma_0$ near the source, diffusivity $D_0$ and quantum capacitance in a form $D_0 C_Q = \sigma_0$ was used.

Another representation of the pinch-off saturation current is $I_{DSAT} = W e n_S(0) v_S$, where the characteristic velocity is defined as

$$v_S = \frac{\sigma_0}{L C_{CH}} = \frac{\mu_0 V_{DSAT}}{2L}. \qquad (9)$$

The current saturation for short-channel FETs (typically $L \leq 0.5$ μm) is bound to the velocity saturation due to scattering on optical phonons. For the velocity saturation $v_{opt}$ it has experimentally and theoretically obtained the relation $v_{opt} = v_0(\hbar\Omega/\varepsilon_F)$, where $\hbar\Omega_{opt} \cong 50$ meV is of order of the optical phonon energy [8]. The channel current saturates due to velocity saturation at $I_{DSAT} = W e n_S(0) v_{opt}$. Note, that for diffusive channel $v_{opt}$ is a maximum velocity of dissipative motion which is in any case less than the speed $v_0$ of ballistic carriers in graphene. One can introduce the dimensionless parameter discriminating the two types of current saturation in FET [9]

$$a = \frac{v_S}{v_{opt}} = \frac{\mu_0 V_{DSAT}}{2 v_{opt} L} = \frac{V_{DSAT}}{V_{D0}}, \qquad (10)$$

where a new characteristic drain voltage is defined

$$V_{D0} \equiv \frac{2 v_{opt} L}{\mu_0}, \qquad V_{D0} \cong 10 \left(\frac{2 v_{opt}}{10^8 cm/s}\right) \left(\frac{L}{1 \mu m}\right) \left(\frac{10^3 cm^2/Vs}{\mu_0}\right) \text{ V.} \qquad (11)$$

When $a \ll 1$ (long channels and thin gate insulators, low carrier densities and mobilities) the electrostatic pinch-off prevails, and if $a \gg 1$ the carrier velocity saturation determines the saturation current of FETs. Thereby the drain current can be rewritten in a unified manner for both cases

$$I_D = W e n_{S0} v_{SAT} \left(1 - \exp\left[-\frac{\mu_0 V_{DS}}{v_{SAT} L}\right]\right) \qquad (12)$$

where $v_{SAT} = \min\{v_{opt}, v_S\}$. A convenient analytical interpolation can be used

$$v_{SAT} = v_{opt} \tanh \frac{v_S}{v_{opt}} = v_{opt} \tanh \frac{V_{DSAT}}{V_{D0}} \qquad (13)$$

which provides convenient analytical description of crossover between two modes of saturation.

Note, that empirical relationships for high-field drift velocity

$$v_{DR}(E) = \frac{\mu_0 E}{\left(1+(\mu_0 E/v_{SAT})^n\right)^{1/n}} \equiv \frac{\mu_0 E}{\left(1+(E/E_{SAT})^n\right)^{1/n}} \quad (14)$$

originating from early work of Thornber [10] and traditionally used in CMOS compact modeling [11] also is nothing but empirical interpolation having besides a significant shortage. This equation does not provide fast saturation and yields only $v_{SAT}/2^{1/n}$ at $E = v_{SAT}/\mu_0$. To remove this shortage for best fitting with experiments a joint interpolation is typically used in CMOS design practice with $E_{SAT} = 2v_{SAT}/\mu_0$ and artificial fitting to obey a formal condition $v(E_{SAT}) = v_{SAT}$. A use of analytic interpolation Eq.13 allows to get rid of piecewise description and senseless fitting parameter $n$.

### B. Unified model for the two saturation modes

Using Eq.12 and 13 a unified relationship for I-V characteristics can be rewritten as

$$I_D = Wen_S v_{opt} \tanh\left(\frac{V_{DSAT}}{V_{D0}}\right)\left(1-\exp\left[-\frac{2V_D}{V_{D0}\tanh(V_{DSAT}/V_{D0})}\right]\right). \quad (15)$$

Notice that near the charge neutrality point when $V_{DSAT} < V_{D0} = 2v_{opt}L/\mu_0$ the "square law" is valid practically at any parameters due to that that electrostatic pinch-off occurs before velocity saturation. This point has confirmed experimentally in Ref.[12]. Further, introducing convenient notation

$$V_{S0} = V_{D0}\tanh\frac{V_{DSAT}}{V_{D0}}, \quad (16)$$

meaning the lesser of $V_{D0}$ or $V_{DSAT}$, the drain current as function of drain-source voltage can be represented by the relation similar to Eq.7

$$I_D = \frac{1}{2}g_{D0}V_{S0}\left(1-\exp\left(-2\frac{V_D}{V_{S0}}\right)\right) \quad (17)$$

where the maximum value of the low-field output conductance is expressed as follows

$$g_{D0} = \frac{W}{L}e\mu_0 n_S(0) = \frac{W}{2L}\mu_0 C_{CH}V_{DSAT} = W C_{CH}v_S. \quad (18)$$

The transit time through the whole channel length for $a > 1$ ($V_{DSAT} < V_{D0}$) was found in Ref.[6] as

$$\tau_{TT} = \int_0^L \frac{dy}{\mu_0(1+\kappa)E(y)} = \frac{L^2}{\mu_0 V_{DSAT}}\coth\left(\frac{V_D}{V_{DSAT}}\right). \quad (19)$$

Velocity saturation requires that the Eq.19 for transit time should be modified as follows

$$\tau_{TT} = \frac{L^2}{\mu_0 V_{S0}}\coth\left(\frac{V_D}{V_{S0}}\right). \quad (20)$$

### 4. Small-signal current parameters model

In this section, the calculation of the current small-signal parameters of GFETs based on modified diffusion-drift model is described. In RF applications the transistor in small-signal amplifiers is operated in the on-state and input small a.c. radio-frequency signals are imposed onto the d.c. gate–source voltage. Here we describe a small-signal equivalent circuit model based on a combination of known physics in the small signal limit and generally common behavior for all field effect type devices.

## A. Input gate admittance and capacitance

The input gate small-signal admittance may be modeled as an inverse sum of the impedances of the gate oxide capacitance and graphene sheet

$$Y_G = \left(\frac{1}{Y_S} + \frac{1}{i\omega C_{ox}}\right)^{-1} = i\omega C_G(\omega) + \text{Re}\, Y_G \qquad (21)$$

Taking into account the interface traps existence the frequency-dependent graphene sheet impedance may be in a single level trap approximation modeled as [13]

$$Y_S(\omega) = i\omega C_Q + i\omega \frac{C_{it}}{1 + i\omega \tau_r} \qquad (22)$$

where $C_{it}$ is the low frequency interface trap capacitance, $\tau_r$ is the characteristic time constant of the interface traps recharging typically smaller than 1 MHz. The frequency-dependent input gate capacitance is determined as

$$C_G(\omega) = \text{Re}\left(Y_G(\omega)/i\omega\right) \qquad (23)$$

For low frequencies $\omega\tau_r \ll 1$ we have the low-frequency gate capacitance

$$C_G = e\left(\frac{\partial N_G}{\partial V_G}\right) = \left(\frac{1}{C_{ox}} + \frac{1}{C_Q + C_{it}}\right)^{-1}, \qquad (24)$$

We are interested here in the high-frequency case $\omega\tau_r \gg 1$ when $Y_S \cong i\omega C_Q$ and the interface traps do not respond to external a.c. gate signal. In this case the high-frequency gate capacitance is frequency and interface traps independent

$$C_G\big|_{\omega\tau_r \gg 1} = C_{CH}\big|_{\omega\tau_r \gg 1} \cong \left(\frac{1}{C_{ox}} + \frac{1}{C_Q}\right)^{-1} \qquad (25)$$

although the presence of the interface traps distorts the C-V characteristics stretching out the C-V curves along the voltage axes. Frequency-dependent response of interface traps yields a peak for $\omega\tau_r \sim 1$ with non-zero $\text{Re}\, Y_G$ carrying information about interface traps parameters (so called conduction method of extraction [13]).

## B. Small-signal response matrices of GFET represented as two-port network

The RF performance of FETs are characterized in terms of small-signal parameters such as internal gate transconductance ($g_m$), the output conductance ($g_D$), and the gate-to-source $C_{GS}$ and the gate-to-drain ($C_{GD}$) capacitances. This is illustrated by a small-signal equivalent circuit in Fig. 1 where $R_D$ and $R_S$ are the drain and the source access resistances, respectively [4, 14].

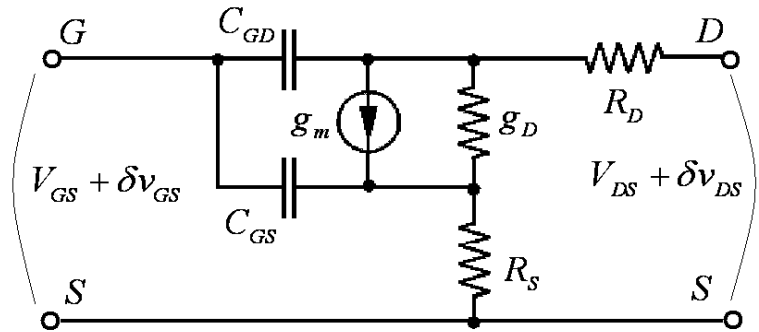

Fig.1. Common source small-signal equivalent circuit of GFET without parasitic capacitances and gate impedance.

Small-signal current equations for two-port pi network

$$\delta i_D = g_m \delta v_{GS} + i\omega C_{GD}(\delta v_{GS} - \delta v_{DS}) + g_D \delta v_{DS} = (g_m + i\omega C_{GD})\delta v_{GS} + (g_D - i\omega C_{GD})\delta v_{DS}$$

$$\delta i_G = i\omega C_{GS} \delta v_{GS} + i\omega C_{GD}(\delta v_{GS} - \delta v_{DS}) = i\omega(C_{GS} + C_{GD})\delta v_{GS} - i\omega C_{GD} \delta v_{DS}$$

may be rewritten in matrix form

$$\begin{pmatrix} \delta i_G \\ \delta i_D \end{pmatrix} = Y \begin{pmatrix} \delta v_{GS} \\ \delta v_{DS} \end{pmatrix}, \qquad (26)$$

where admittance matrix is defined as

$$Y \equiv \begin{pmatrix} i\omega(C_{GS} + C_{GD}) & -i\omega C_{GD} \\ g_m + i\omega C_{GD} & g_D - i\omega C_{GD} \end{pmatrix}. \qquad (27)$$

Inversing the admittance matrix we get the intrinsic impedance matrix $Z = Y^{-1}$ with the components

$$z_{11} = \frac{g_D - i\omega C_{GD}}{-i\omega(C_{GG}g_D + C_{GD}g_m - i\omega C_{GS}C_{GD})}, \qquad (28a)$$

$$z_{12} = \frac{C_{GD}}{C_{GG}g_D + C_{GD}g_m - i\omega C_{GS}C_{GD}}, \qquad (28b)$$

$$z_{21} = \frac{g_m + i\omega C_{GD}}{-i\omega(C_{GG}g_D + C_{GD}g_m - i\omega C_{GS}C_{GD})}, \qquad (28c)$$

$$z_{22} = \frac{C_{GG}}{C_{GG}g_D + C_{GD}g_m - i\omega C_{GS}C_{GD}}. \qquad (28d)$$

The components of admittance matrix describe high-frequency small-signal response of field-effect transistors.

### C. Intrinsic output conductance and transconductance

Ignoring many complications one can conclude that current-voltage characteristics with saturation may easily parameterized by the two parameters: the output conductance and the saturation voltage. The drain conductance as function of the node biases (closely connected with low-field conductance $g_{D0}$) can be calculated using Eq.17 as a partial derivative with fixed $V_{GS}$

$$g_D = \left(\frac{\partial I_D}{\partial V_{DS}}\right)_{V_{GS}} = g_{D0} \exp\left(-\frac{2V_{DS}}{V_{S0}}\right), \qquad (29)$$

where $g_{D0}$ is given by Eq.18.

One of the most important small-signal parameter for high-frequency performance prediction is the intrinsic gate transconductance $g_m$. Transconductance depends generally on microscopic mobility slightly varying with the gate voltage the underlying mechanism and quantitative description of that has not been yet understood and developed in details. Omitting here this point the microscopic mobility will be considered as to be independent on the gate bias in this paper. Exact view of relation the intrinsic transconductance for arbitrary value of the parameter *a* depends on the choice of approximation for current and has awkward form. We will use here a convenient approximation

$$g_m \cong \frac{W}{2L}\mu_0 C_{CH} V_{S0}\left(1 - \exp\left(-\frac{2V_{DS}}{V_{S0}}\right)\right) \qquad (30)$$

The transconductance the $g_m$ increases linearly with $V_{DS}$ up to saturation on maximum level

$$g_m^{(\max)} \cong \frac{W}{2L}\mu_0 C_{CH} V_{S0} = \begin{cases} \frac{W}{2L}\mu_0 C_{CH} V_{DSAT} = \frac{W}{L}\mu_0 e n_S(0) = W C_{CH} v_S, & V_{D0} > V_{DSAT} \\ \frac{W}{2L}\mu_0 C_{CH} V_{D0} = W C_{CH} v_{opt}, & V_{DSAT} > V_{D0}. \end{cases} \qquad (31)$$

## D. Role of the contact resistances

Significant limitation of state-of-the-art GFETs is the substantial access resistance due to the significant gaps between the source-gate and gate-drain electrodes, where a large portion of the graphene channel in the gap area is not gated. Access and parasitic contact resistances can significantly degrade performance characteristics of GFETs. Unfortunately the typical values of state-of-the-art graphene-metal contacts may be as high as hundreds of Ohm×μm and larger [15]. The voltage drops on the source and the drain resistances ($R_S$ and $R_D$) stipulate that internal node voltages ($V_{GS}^{int}$ and $V_{DS}^{int}$) immediately governing the drain current may be sufficiently smaller than the voltages applied to external contacts ($V_{GS}^{ext}$ and $V_{DS}^{ext}$). For voltage-independent contact resistances the elementary Kirchhoff laws yield $V_{GS}^{int} = V_{GS}^{ext} - I_D\left(V_{GS}^{int}, V_{DS}^{int}\right) R_S$, $V_{DS}^{int} = V_{DS}^{ext} - I_D\left(V_{GS}^{int}, V_{DS}^{int}\right)(R_S + R_D)$. Taking the derivatives of the drain current as function of two internal voltages $g_m^{ext} = g_m^{int}\left(\partial V_{GS}^{int}/\partial V_{GS}^{ext}\right) + g_D^{int}\left(\partial V_{DS}^{int}/\partial V_{DS}^{ext}\right)$ one can obtained useful relationships connecting intrinsic and extrinsic steady-state (DC) parameters (see, for example [16]

$$g_m^{ext} = \frac{\partial I_D}{\partial V_{GS}^{ext}} = \frac{g_m^{int}}{1 + g_m^{int} R_S + g_D^{int}(R_S + R_D)}, \quad g_D^{ext} = \frac{\partial I_D}{\partial V_{DS}^{ext}} = \frac{g_D^{int}}{1 + g_m^{int} R_S + g_D^{int}(R_S + R_D)}. \quad (32)$$

The contact and access resistances may be fundamentally limiting predicted gain and THz capabilities of GFETs, since poor contacts can significantly decrease $g_m$ and $f_T$. [17, 18]. The formation of the self-aligned source and drain electrodes allows precise positioning of the source–drain edges with the gate edges, and thus substantially reduces the access resistance and improves the performance of the graphene transistor [5].

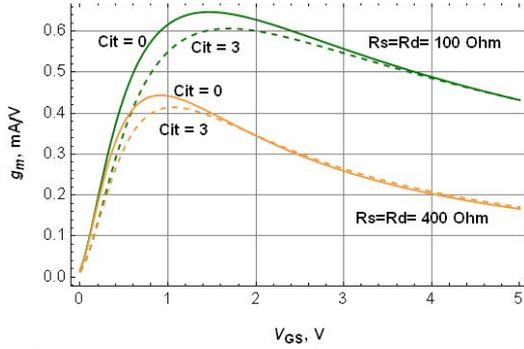
Fig. 2a. Simulated extrinsic transconductance as functions of $V_{GS}$ at $V_{DS} = 1$ V for different contact resistances $R_S = R_D = 400$ Ω×μm (lower curves) and $R_S = R_D = 100$ Ω×μm (upper curve) for zero interface trap density (solid lines) and $C_{it} = 3$ $fF/\mu m^2$ (dashed lines); $W/L = 1\mu m/0.5$ μm, $\mu_0 = 2000$ cm$^2$/V-s; $d_{ox} = 10$ nm, $\varepsilon_{ox}=4$, T=300 K, $v_{opt} = 0.5\ v_0$

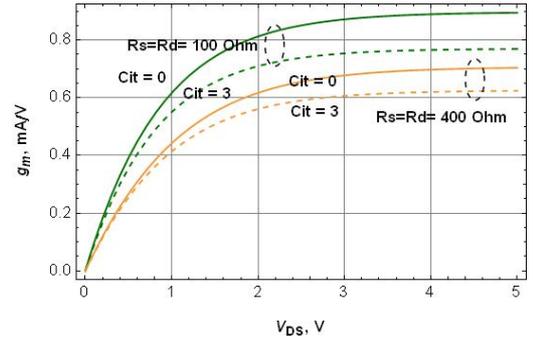
Fig. 2b. Simulated transconductance as functions of $V_{DS}$ at $V_{GS} = 1$ V for different contact resistances $R_S = R_D = 400$ Ω×μm (lower curves, brown lines online) and $R_S = R_D = 100$ Ω×μm (upper curves, green lines online) for zero interface trap density (solid lines) and $C_{it} = 3$ $fF/\mu m^2$ (dashed lines); $W/L = 1\mu m/0.5$ μm, $\mu_0 = 2000$ cm$^2$/V-s; $d_{ox} = 10$ nm, $\varepsilon_{ox}=4$, T=300 K, $v_{opt} = 0.5\ v_0$

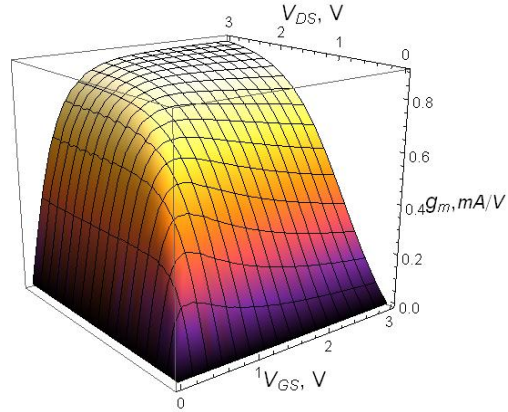
Fig. 2c. Simulated extrinsic transconductance as functions of $V_{GS}$ and $V_{DS}$ at $W/L = 1\mu m/0.2$ μm, $R_S = R_D = 500$ Ω×μm, $\mu_0 = 2000$ cm$^2$/V-s; $d_{ox} = 10$ nm, $\varepsilon_{ox}=4$, T=300 K, $v_{opt} = 0.5\ v_0$, $C_{it} = 0$

Figs.2 display simulated transconductance as function of $V_{GS}$ and $V_{DS}$ at different parasitic resistances and interface trap capacitances. Notice that although the interface traps do not respond to input high-frequency small a.c. signals they respond to relatively slow large-signal d.c. $V_{GS}$ variation.

## 5. Small-signal capacitance model

To obtain mutual capacitances between the gate, source and drain one should derive an explicit dependence of the full channel charge $Q_C(V_{GS}, V_{DS})$ as explicit function of the node voltages. It can only be done using a detailed model of charge and potential distribution in Ref. [6]. The Eq.1b for the chemical potential distribution can rewritten as follows

$$\varepsilon_F(x) = \varepsilon_F(0) + \frac{\varepsilon_F}{2} \ln\left[1 - \frac{x}{L}\left[1 - \exp\left(-\frac{2V_{DS}}{V_{DSAT}}\right)\right]\right], \qquad (33)$$

where $\varepsilon_F(0)$ is the Fermi energy near the drain imposed by the gate-source voltage. The onset of saturation at $V_{DS} = V_{DSAT}$ corresponds exactly to the zero Fermi energy near the drain $\varepsilon_F(L) = 0$. Once $V_{DS}$ exceeds $V_{DSAT}$ the conduction type at the drain end of the channel switches from n-type to p-type and the chemical potential near the drain becomes the negative. The region of positive charge creeps into the channel as $V_{DS}$ increases and its front is determined by the condition $\varepsilon_F(x_0) = 0$

$$x_0(V_{DS}) = \frac{1 - e^{-2}}{1 - e^{-\frac{2V_{DS}}{V_{DSAT}}}} L, \qquad V_{DS} \geq V_{DSAT}. \qquad (34)$$

### A. Channel charge as function of the node biases

To obtain the full channel charge $Q_C(V_{GS}, V_{DS})$ one should to integrate the channel charge density distribution over the channel length $L$ assuming validity of the gradual channel approximation (i.e. the condition of planar electric neutrality along the channel).

$$\frac{Q_C(V_{GS}, V_{DS})}{Q_{C0}(V_{GS})} = \int_0^L \left(\frac{\varepsilon_F(x)}{\varepsilon_F(0)}\right)^2 \frac{dx}{L} \qquad V_{DS} < V_{DSAT} \qquad (35a)$$

$$\frac{Q_C(V_{GS}, V_{DS})}{Q_{C0}(V_{GS})} = \int_0^{x_0(V_{DS})} \left(\frac{\varepsilon_F(x)}{\varepsilon_F(0)}\right)^2 \frac{dx}{L} - \int_{x_0(V_{DS})}^L \left(\frac{\varepsilon_F(x)}{\varepsilon_F(0)}\right)^2 \frac{dx}{L} \qquad V_{DS} > V_{DSAT} \qquad (35b)$$

where $Q_{C0} = Q_C(V_{GS}, V_{DS}=0)$ and the first and the second terms in Eq.35b correspond to the full charge of the electron and the hole parts of the channel correspondingly. Performing integration with an explicit dependence $\varepsilon_F(x)$ from Eq.1 we have found the channel charge as function of dimensionless variable $s \equiv V_{DS}/V_{DSAT}$ introduced for brevity

$$Q_C = Q_{C0}(V_{GS}) F(s) \qquad (36a)$$

where dimensionless $F(s)$ function is defined as follows

$$F(s) = \begin{cases} \dfrac{1}{2}\left[1 + (1-s)s(\coth s - 1)\right], & s \leq 1 \\ \dfrac{1}{4}\left[\overbrace{(1 + \coth s)(1 - e^{-2})}^{\text{electrons}} - \overbrace{\dfrac{e^{s-2} - e^{-s}(1 - 2(1-s)s)}{\sinh s}}^{\text{holes}}\right], & s > 1 \end{cases} \qquad (36b)$$

It is useful to determine the derivative of the $F(s)$

$$-\frac{dF(s)}{ds} = \begin{cases} \frac{1}{2}\left[(2s-1)(\coth s - 1) - \frac{(s-1)s}{\sinh^2 s}\right], & s \leq 1 \\ \frac{e^{-s}}{2\sinh s}\left[(s-2)(s-1) - e^{-2} + \left((s(s-1)+1) - e^{-2}\right)\coth s\right], & s > 1. \end{cases} \qquad (37)$$

Despite of its piecewise character the universal $F(s)$ dependence behaves as a smooth function of its variable (see Fig.3)

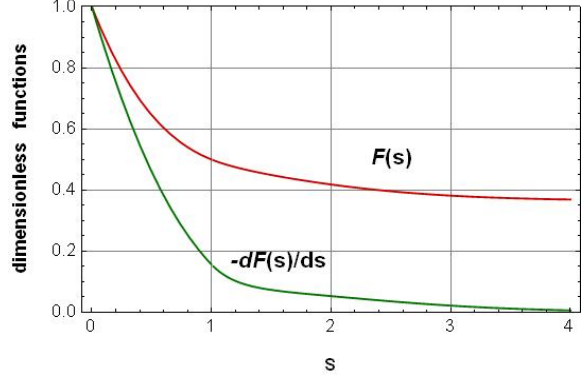

Fig. 3. Universal curves $F(s)$ and its derivative.

### B. Gate-source and gate-drain capacitances

Neglecting charged oxide defects (interface traps) we have equality of the gate and the channel capacitances $C_G = C_C = (\partial Q_C/\partial V_{GS}) = (e\partial N_G/\partial V_{GS})$. For the common source circuit the total gate capacitances $C_{GG}$ at finite drain-source $V_{DS}$ is defined as

$$C_{GG} = C_{GS} + C_{GD} = \left(\frac{\partial Q_C}{\partial V_{GS}}\right)_{V_{DS}}, \qquad (36)$$

and taking into account $V_{GD} = V_{GS} - V_{DS}$ one obtains the full gate-drain and the gate-source capacitances $C_{GD}$ and $C_{GS}$ as functions of the gate and the drain voltages

$$C_{GD} = \left(\frac{\partial Q_C}{\partial V_{GD}}\right)_{V_{GS}} = -\left(\frac{\partial Q_C}{\partial V_{DS}}\right)_{V_{GS}}, \qquad C_{GS} = \left(\frac{\partial Q_C}{\partial V_{GS}}\right)_{V_{DS}} + \left(\frac{\partial Q_C}{\partial V_{DS}}\right)_{V_{GS}}. \qquad (37)$$

The magnitude of the ratio $C_{GS}$ to $C_{GD}$ characterizes an extent of the channel charge control by the gate and absence of the direct coupling capacitance between the gate and drain nodes [19]. Taking into account Eq.7 the direct differentiations yield

$$C_{GD} = \frac{Q_{C0}(V_{GS})}{V_{DSAT}}\left(-\frac{dF(s)}{ds}\right) = \frac{WLC_{CH}}{2}\left(-\frac{dF(s)}{ds}\right), \qquad (38a)$$

$$C_{GG} = WLC_{CH}\left[F(s) + \frac{s}{2}\left(-\frac{dF(s)}{ds}\right)\left(\frac{\partial V_{DSAT}}{\partial V_{GS}}\right)\right], \qquad (38b)$$

$$C_{GS} = WLC_{CH}\left[F(s) + \frac{1}{2}\left(-\frac{dF(s)}{ds}\right)\left(s\left(\frac{\partial V_{DSAT}}{\partial V_{GS}}\right) - 1\right)\right]. \qquad (38c)$$

The relationship

$$\frac{\partial V_{DSAT}}{\partial V_{GS}} = 1 + \frac{C_{CH}}{C_{ox}},$$

following from Eq.4, accomplishes a closed set of explicit equations for analytical calculation of small-signal capacitance characteristics. Simulated small-signal capacitance dependencies on $V_{DS}$ and/or $V_{GS}$ are depicted in Figs.4-6.

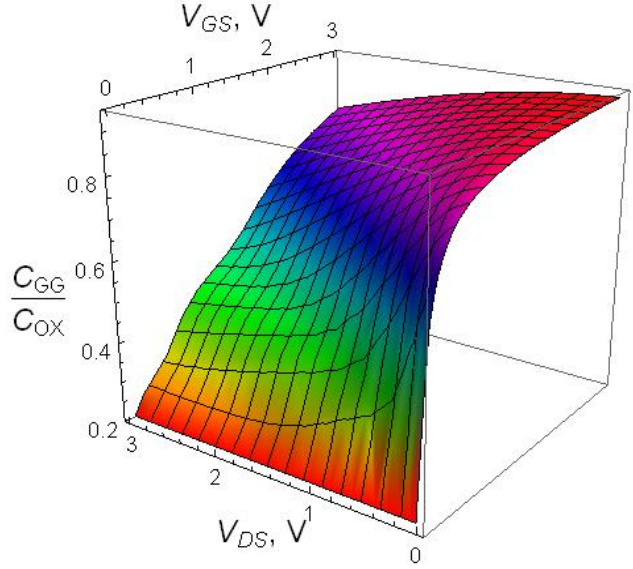

Fig.4. Simulated gate capacitance as functions of gate-source and drain source voltages.

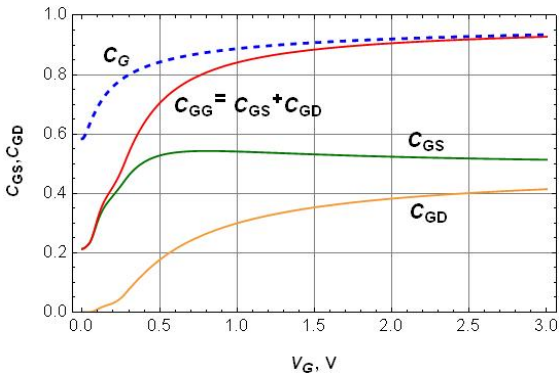

Fig.5a. Simulated small-signal capacitances normalized to $C_{ox}$ as functions of $V_{GS}$ ($V_{NP} = 0$) at $V_{DS} = 0.5$ V; $d_{ox} = 10$ nm, $\varepsilon_{ox}=4$, (room temperature)

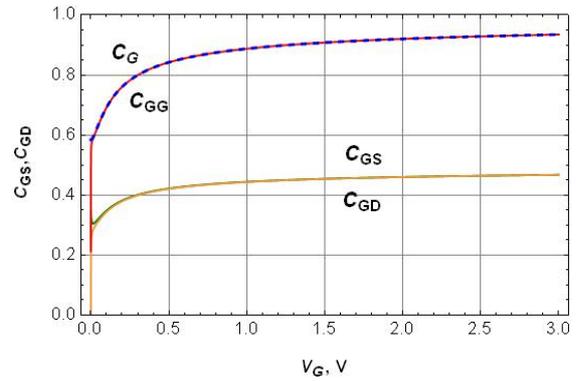

Fig.5b. The same curves calculated at $V_{DS} = 0.001$ V; $C_{GG} = C_G = 2\, C_{GD} = 2\, C_{GS}$ as expected

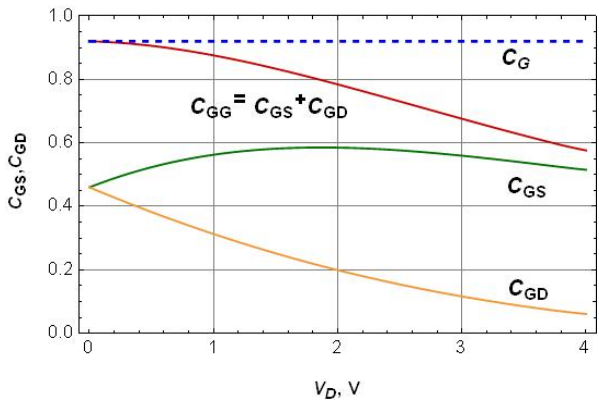

Fig. 6a. Simulated small-signal capacitances normalized to $C_{ox}$ as functions of $V_{DS}$ at $V_{GS} = 2$ V; $V_{NP} = 0$; $d_{ox} = 10$ nm, $\varepsilon_{ox}=4$, (room temperature)

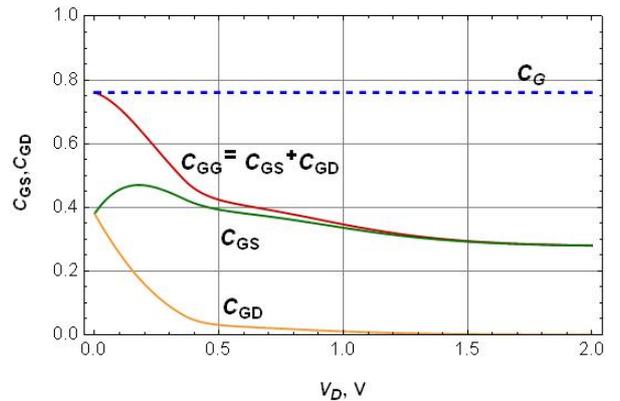

Fig. 6b. The same curves calculated at $V_{GS} = 0.2$ V.

The equations can be easily generalized for non-zero interface trap density case noticing that the gate and the channel capacitances are connected with each other as [6]

$$\frac{C_G}{C_{CH}} = 1 + \frac{C_{it}}{C_Q}$$

Of course, there are unavoidably additional parasitic reciprocal capacitive electrostatic couplings between the gate and source/drain electrodes which were not considered above. These parasitic elements limit the drive capability and switching speed of the device and should be minimized but cannot be eliminated entirely.

## 6. Frequency-dependent current gain and cutoff frequency

This section is devoted to modeling of the complex current gain $h_{21}$ and the cutoff frequency $f_T$ which is one of the main figures-of-merit of RF transistors.

### A. Intrinsic current gain and cutoff frequency

Neglecting temporarily the problem of parasitic capacitances and series resistances in the source-drain circuit the intrinsic short-circuit current gain can be written as

$$h_{21}^{(int)}(\omega) = \left(\frac{\partial I_D}{\partial I_G}\right)_{V_{GS}} = -\frac{(\partial V_{GS}/\partial I_G)_{I_D}}{(\partial V_{GS}/\partial I_D)_{I_G}} = -\frac{z_{21}}{z_{22}} = \frac{g_m + i\omega C_{GD}}{i\omega C_{GG}}. \quad (39)$$

The cut-off frequency $f_T$ defined as the frequency at which the gain becomes unity is the most widely used figure-of-merit for RF devices and is, in effect, the highest frequency a FET is useful in RF applications. The condition $|h_{21}(\omega_T)| = 1$ yields

$$2\pi f_T = \frac{g_m}{\left(C_{GG}^2 - C_{GD}^2\right)^{1/2}} = \frac{g_m}{C_{GG}\left(1 - \eta^2\right)^{1/2}}$$

where $\eta = C_{GD}/C_{GG}$. Omitting for brevity the factor $\eta$ typically small for large $V_{DS}$ and taking into account Eqs.30 and 38 the cut-off frequency may be represented as follows

$$2\pi f_T = \frac{\mu_0 V_{S0}}{2L^2} \frac{1 - \exp(-2V_{DS}/V_{S0})}{F(s) + \frac{1}{2}\left(\frac{V_{DS}}{V_{DSAT}}\right)\left(1 + \frac{C_{CH}}{C_{ox}}\right)\left(-F'\left(\frac{V_{DS}}{V_{DSAT}}\right)\right)} \quad (40)$$

Simple analysis yields that low $V_{DS}$ we have $f_T \propto \mu_0 V_{DS}/L^2$ and in the current saturation mode ($V_{DS} > V_{S0}$)

$$2\pi f_T \cong \frac{\mu_0 V_{S0}}{2L^2} = \begin{cases} \frac{\mu_0 V_{D0}}{2L^2} = \frac{v_{opt}}{L}, & V_{DS} > V_{DSAT} > V_{D0} \\ \frac{\mu_0 V_{DSAT}}{2L^2}, & V_{DS} > V_{D0} > V_{DSAT}. \end{cases} \quad (41)$$

Naturally, the intrinsic cutoff frequency in any case is determined by the carrier transit time through the channel. Both $L^{-1}$ in short-channel GFETs [20, 21] and $L^{-2}$ [8] has been observed experimentally depending on operation mode.

### B. Extrinsic current gain and cut-off frequency

Extrinsic short circuit small-signal current gain depends additionally on contact drain and source resistance ($R_{SD} = R_S + R_D$). Modifying Eq. 39 we have

$$h_{21}^{ext}(\omega) = -\frac{z_{21} + R_S}{z_{22} + R_{SD}} \quad (42)$$

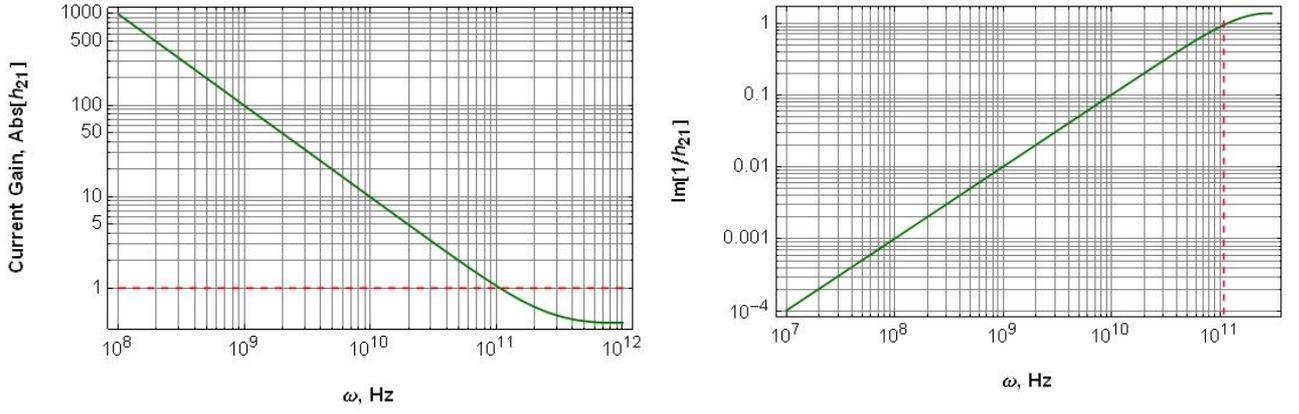

Fig.7. Absolute value of complex current gain Abs[$h_{21}$] and the imagine part of $-1/h_{21}$ simulated as functions of frequency demonstrating typical slopes 20 dB/decade at $f < f_T$. Intersections with dashed lines correspond to the cutoff frequency $f_T$ (~100 GHz) Simulations performed for $W=L=1\mu m$; $\mu_0 = 2000$ cm$^2$/V-s; $V_{DS} = V_{GS} = 1$ V, $d_{ox} = 10$ nm, $\varepsilon_{ox}=4$, T=300 K, $C_{it} = 3$ $fF/\mu m^2$, $R_S = R_D = 1$ k$\Omega$, $d_{ox} =10$ nm, $v_{opt} = 0.5$ $v_0$.

Fig.7 shows simulated dependencies of Abs[$h_{21}$] and Im [$1/h_{21}$] as functions of frequency exhibiting "ideal" slope -20 dB/decade.

Equating to unity the extrinsic short-circuit current gain modified for non-zero $R_{SD}$

$$\left| h_{21}^{ext} \left( \omega_T^{ext} \right) \right| = \left| \frac{z_{21} + R_S}{z_{22} + R_{SD}} \right| = 1 \qquad (42)$$

and neglecting $\omega C_{GD} \ll g_m$ and $\omega C_{GS} C_{GD} \ll C_{GG} g_D + C_{GD} g_m$ one can found the extrinsic cut-off frequency

$$f_T \cong \frac{g_m}{2\pi} \frac{1}{\left[ \left( C_{GG}(1+g_D R_{SD}) + C_{GD} g_m R_{SD} \right)^2 - \left( C_{GG} g_D + C_{GD} g_m \right)^2 R_S^2 \right]^{1/2}}. \qquad (44)$$

In practice, due to unavoidable parasitic capacitance presence the cut-off frequency $f_T$ might be significantly below its theoretical maximum value. We modeled here this modifying Eq.43 as follows $C_{GG} \rightarrow C_{GG} + par\, C_{ox} WL$ and $C_{GD} \rightarrow C_{GD} + par\, C_{ox} WL$, where *par* is dimensionless factor characterizing a fraction of parasitic capacitance.

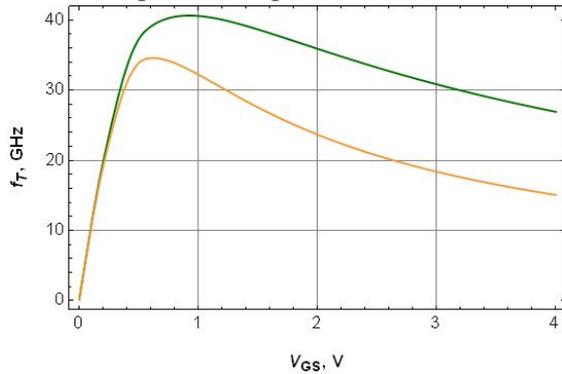
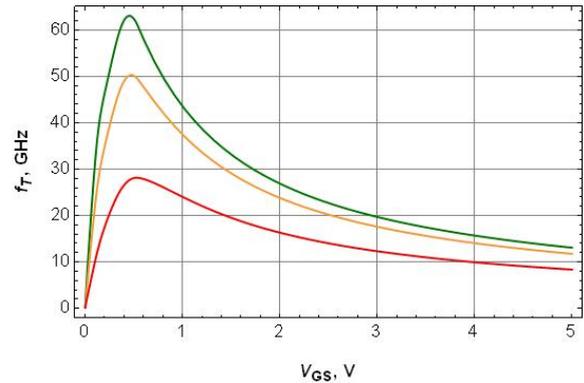

Fig. 8. Simulated cut-off frequency $f_T$ as functions of $V_{GS}$ at $V_{DS}=1$ V for different contact resistances $R_S = R_D=400$ $\Omega\times\mu m$ (lower curve) and $R_S = R_D=100$ $\Omega\times\mu m$ (upper curve) $W/L = 1\mu m/0.5$ $\mu m$, $\mu_0 = 2000$ cm$^2$/V-s; $d_{ox} =10$ nm, $\varepsilon_{ox}=4$, T=300 K and $C_{it} = 0$ $fF/\mu m^2$, $v_{opt} = 0.5$ $v_0$, *par*=0.5

Fig.9. Simulated cut-off frequency $f_T$ as function of $V_{GS}$ at $V_{DS}=1$ V for different parasitic capacitances *par*=0.5 (lower curve), *par*=0 (upper) and *par*=0.1 (intermediate); $W/L = 1\mu m/0.5$ $\mu m$, $\mu_0 = 2000$ cm$^2$/V-s; $d_{ox} =10$ nm, $\varepsilon_{ox}=4$, T=300 K, $v_{opt} = 0.5$ $v_0$, $R_S = R_D=400$ $\Omega\times\mu m$, $C_{it} = 0$ $fF/\mu m^2$

As can be seen from the Figs.8-9, the cut-off frequency $f_T$ can be maximized through minimization of parasitic capacitances and resistances.

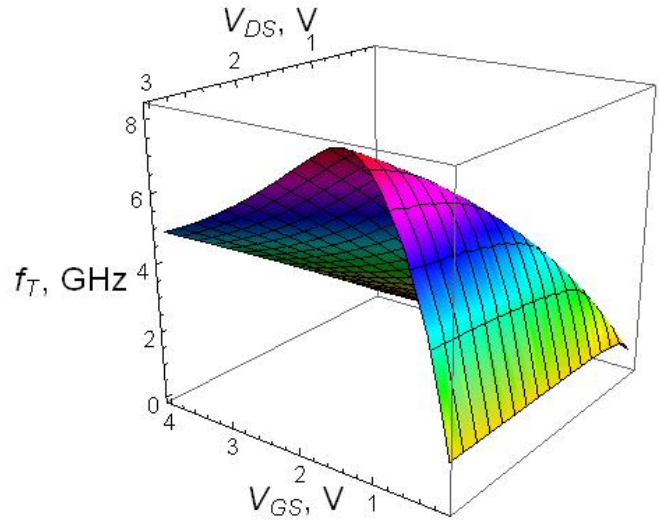

Fig.10. Simulated cut-off frequency $f_T$ as simultaneous function of $V_{GS}$ and $V_{DS}$ calculated for and $C_{it} = 0$ $fF/\mu m^2$, $W/L = 1\mu m/1$ μm, $\mu_0 = 1000$ cm$^2$/V-s; $d_{ox} = 10$ nm, $\varepsilon_{ox}=4$, T=300 K, $v_{opt} = 0.5$ $v_0$, $par=0.5$, $R_S = R_D=1$ kΩ×μm.

Fig.10 displays 3D plot for simulated dependence of extrinsic cut-off frequency as function gate-source and drain-source voltages. Characteristic peak of the $f_T$ dependence on $V_{GS}$ [22] is explained by presence of the peak in $g_m(V_{GS})$ dependence (see Fig.2a) appearing mainly because of the parasitic resistances presence.

# References

[1] K. S. Novoselov et al., "Electric Field Effect in Atomically Thin Carbon Films," Science, vol. 306, 2004, p. 666.

[2] M. Lemme et al., "A Graphene Field-Effect Device," IEEE Electron Device Letters, vol. 28, 2007, pp. 282–84.

[3] T. Palacios, A. Hsu, and H. Wang, "Applications of Graphene Devices in RF Communications," IEEE Communications Magazine, No.6, pp. 122-128, June 2010.

[4] F. Schwierz, "Graphene Transistors," Nature Nanotechnology, 30 May 2010 | doi: 10.1038/nnano.2010.89

[5] Lei Liao et al., "High-speed graphene transistors with a self-aligned nanowire gate," doi:10.1038/nature09405, 2010.

[6] G. I. Zebrev, "Graphene Field Effect Transistors: Diffusion-Drift Theory", a chapter in "*Physics and Applications of Graphene – Theory*", Ed. by S. Mikhailov, Intech, 2011.

[7] S.M. Sze, K.K. Ng, "*Physics of Semiconductor Devices*," 3$^{rd}$ edition, Wiley-Interscience, 2007.

[8] I. Meric et al. "RF performance of top-gated, zero-bandgap graphene field-effect transistors," IEDM, 2008

[9] G.I. Zebrev, Fiz. Tekh. Poluprovodn. (Sov. Phys. Semicond.) "Current-voltage characteristics of a metal-oxide-semiconductor transistor calculated allowing for the dependence of the mobility on a longitudinal electric field," V. 26, No.1, (1992).

[10] K.K. Thornber, "Relation of drift velocity to low-field and high-field saturation velocity," J. Appl. Phys. 1980, 51, 2127.

[11] Y. Cheng, C. Hu, "*MOSFET Modeling & BSIM3 User's Guide*," Kluwer Academic Publishers, 2002.

[12] J. S. Moon, D. Curtis, D. Zehnder, S. Kim, D. K. Gaskill, G. G. Jernigan, R. L. Myers-Ward, C. R. Eddy, Jr., P. M. Campbell, K.-M. Lee, and P. Asbeck, "Low-Phase-Noise Graphene FETs in Ambipolar RF Applications," IEEE Electron Device Letters, Vol. 32, No. 3, March 2011"

[13] E.H. Nicollian, J.R Brews, 1982, *MOS (Metal Oxide Semiconductor) Physics and Technology*, Bell Laboratories, Murray Hill, USA

[14] F. Schwierz, J.J. Liou, "RF transistors: Recent developments and roadmap toward terahertz applications," Solid. State Electronics, 51, 1079–1091(2007)

[15] S. Russo et al., "Contact resistance in graphene-based devices", Physica E, Volume 42, Issue 4, p. 677-679, 2010

[16] M.S. Shur, "*Physics of Semiconductor Devices*," Prentice-Hall International, Inc., 1990.

[17] "Impact of contact resistance on the transconductance and linearity of graphene transistors", Appl. Phys. Lett. 98, 183505 (2011)

[18] X. Yang, G. Liu, M. Rostami, A. Balandin, and K. Mohanram "Graphene Ambipolar Multiplier Phase Detector" IEEE Trans. Electron Devices (in press).

[19] J.-P. Raskin, "SOI Technology: An Opportunity for RF Designers?" Journal of Communications and Information Technology, No.4, 2009.

[20] Chun-Yung Sung, "Graphene nanoelectronics," ISDRS, Dec. 2009

[21 ] Y.Wu, Y.-M. Lin et al. "High-frequency, scaled graphene transistors on diamond-like carbon," Nature, 472 (7341), 74-78 (2011)

[22] Yu-Ming Lin, Keith A. Jenkins, Alberto Valdes-Garcia, Joshua P. Small, Damon B. Farmer, and Phaedon Avouris "Operation of Graphene Transistors at Gigahertz Frequencies", Nano Letters, V.9, No.1, 422-426, 2009.